\documentclass[10pt,a4paper]{article}
\usepackage[dvips]{color}
\usepackage{epsfig}
\usepackage{amsmath}
\usepackage{graphicx}

\begin{document}
\title{\bf Phenomenological Fluids from Interacting Tachyonic Scalar Fields}
\author{{J. Sadeghi$^{a}$ \thanks{Email: pouriya@ipm.ir},\hspace{1mm} M. Khurshudyan$^{b, c, d}$ \thanks{Email: martiros.khurshudyan@nano.cnr.it},
\hspace{1mm} M. Hakobyan$^{e,d}$
\thanks{Email: margarit@mail.yerphi.am}\hspace{1mm} and H.
Farahani$^{a}$ \thanks{Email:
h.farahani@umz.ac.ir}}\\
$^{a}${\small {\em Department of Physics, Mazandaran University, Babolsar, Iran}}\\
{\small {\em P .O .Box 47416-95447, Babolsar, Iran}}\\
$^{b}${\small {\em CNR NANO Research Center S3, Via Campi 213a, 41125 Modena MO}}\\
$^{c}${\small {\em Dipartimento di Scienze Fisiche, Informatiche e Matematiche,}}\\
{\small {\em Universita degli Studi di Modena e Reggio Emilia, Modena, Italy}}\\
$^{d}${\small {\em Department of Theoretical Physics, Yerevan State
University, 1 Alex Manookian, 0025, Yerevan, Armenia}}\\
$^{e}${\small {\em A.I. Alikhanyan National Science Laboratory,
Alikhanian Brothers St., Yerevan, Armenia} } \\
$^{d}${\small {\em Department of Nuclear Physics, Yerevan State University, Yerevan, Armenia}}\\}  \maketitle

\begin{abstract}
In this paper we are interested to consider mathematical ways to
obtain different phenomenological fluids from two-component
Tachyonic scalar fields. We consider interaction between components
and investigate problem numerically. Statefinder diagnostics and
validity of the generalized second law of thermodynamics performed
and checked. We suppose that our Universe bounded by Hubble
horizon.
\end{abstract}

\section{Introduction}
Modern cosmology faced with a problem when a set of observational
data reveal the following picture of our Universe called modern era
in theoretical cosmology. Observations of high redshift type SNIa
supernovae [1-3] reveal the accelerating expansion of our Universe.
Then, other series of observations like to investigation of surveys
of clusters of galaxies show that the density of matter is very much
less than critical density [4], observations of Cosmic Microwave
Background (CMB) anisotropy indicate that the Universe is flat and
the total energy density is very close to the critical
$\Omega_{\small{tot}} \simeq1$ [5]. Explanation of accelerated
expansion of our Universe takes two different ways and now they are
developing and evaluating as different approaches, however there is
not any natural restriction of a possibilities of recombination of
two approaches in one single approach. In that case we believe that
joined approach will be more sufficient and rich with new and
interesting physics. To explain recent observational data, which
reveals accelerated expansion character of the Universe, several
models were proposed. One of the possible scenarios (general
relativity framework) is the existence of a dark energy with
negative pressure and positive energy density giving an acceleration
to the expansion [6, 7]. Now, astronomical studies and observational
data of WMAP confirm that dark energy occupies about 73$\% $ of the
energy of our Universe. Other component, dark matter occupies about
23$\%$, and usual baryonic matter
occupies about 4$\%$.\\
There are several models to describe dark energy such as the
cosmological constant and its generalizations [8], or Chaplygin gas
and its generalizations [9-15].  Also there are other candidates for
the dark energy which are scalar-field dark energy models. A
quintessence field [16] is a scalar field with standard kinetic
term, which minimally coupled to gravity. In that case the action
has a wrong sign kinetic term and the scalar field is called phantom
[17]. Combination of the quintessence and the phantom is known as
the quintom model [18]. Extension of kinetic term in Lagrangian
yields to a more general frame work on field theoretic dark energy,
which is called k-essense [19, 20]. A singular limit of k-essense is
called Cuscuton model [21]. This model has an infinite propagating
speed for linear perturbations, however causality is still valid.
The most general form for a scalar field with second order equation
of motion is the Galileon field which also could behaves as dark
energy [22].\\
Among different viewpoints concerning to the nature of the dark
component of the Universe, in this article, we accept that it could
be described by a scalar field and we choose Tachyon as a scalar
field. This paper is organized as follows. After introduction, in
the next section, we consider two-component Tachyonic fluid and give
motivation of our work. In section 3 we write field equations
include an interaction between components. Then check validity of
the Generalized Second Law (GSL) of thermodynamics for this setting
in section 4. In section 5 the statefinder diagnostics is performed
as well. In section 6 we give numerical analysis of the differential
equations which obtained in previous sections. Finally in section 7
we give conclusion and discuss about results.
\section{Tachyonic fluid}
Tachyonic fluid described by the following relativistic Lagrangian
[23],
\begin{equation}\label{eq:tach lag}
L_{TF}=-V(\phi)\sqrt{1-\partial_{i}\phi\partial^{i}\phi}.
\end{equation}
Therefore, the stress energy tensor is given by,
\begin{equation}\label{eq:energy tensor}
T^{ij}=\frac{\partial L_{TF}}{\partial
(\partial_{i}\phi)}\partial^{j}\phi-g^{ij}L_{TF},
\end{equation}
which yields to the following energy density and pressure,
\begin{equation}\label{eq:tachyonic density}
\rho=\frac{V(\phi)}{\sqrt{1- \partial_{i}\phi \partial^{i}\phi}},
\end{equation}
\begin{equation}\label{eq:tachyonic pressure}
P=-V(\phi)\sqrt{1- \partial_{i}\phi \partial^{i}\phi},
\end{equation}
respectively. Our next step is to present the equations (3) and (4)
as the following,
\begin{eqnarray}\label{eq:presdens}
\rho&=&\rho_{1}+\rho_{2},\nonumber\\
P&=&P_{1}+P_{2},
\end{eqnarray}
with the following components,
\begin{eqnarray}\label{matter}
\rho_{1}&=&\frac{V(\phi)\partial_{i}\phi \partial^{i}\phi}{\sqrt{1-
\partial_{i}\phi \partial^{i}\phi}} + \alpha F(H),\nonumber\\
P_{1}&=& \beta K(H),
\end{eqnarray}
and,
\begin{eqnarray}\label{eq:cosconst}
\rho_{2}&=& V(\phi)\sqrt{1- \partial_{i}\phi
\partial^{i}\phi}-\alpha F(H),\nonumber\\
P_{2}&=&-V(\phi)\sqrt{1-
\partial_{i}\phi
\partial^{i}\phi}-\beta K(H),
\end{eqnarray}
where $F(H)$ and $K(H)$ are arbitrary function of $H$. Under such
splitting we have two-component fluid. First of them describes by
EoS (Equation of State) parameter $\omega_{1}$, which is given by,
\begin{equation}\label{eq:omega1}
\omega_{1}=\frac{\beta K(H) \sqrt{1- \partial_{i}\phi \partial^{i}\phi}}{V(\phi)\partial_{i}\phi \partial^{i}\phi+\alpha
F(H)\sqrt{1- \partial_{i}\phi \partial^{i}\phi}}.
\end{equation}
EoS parameter for the second component reads as,
\begin{equation}\label{eq:omega2}
\omega_{2}=\frac{-V(\phi)\sqrt{1- \partial_{i}\phi
\partial^{i}\phi}-\beta K(H)}{V(\phi)\sqrt{1- \partial_{i}\phi
\partial^{i}\phi}-\alpha F(H)},
\end{equation}
while the total EoS of fluid is
\begin{equation}\label{eq:omegatotal}
\omega= -(1-\partial_{i}\phi \partial^{i}\phi).
\end{equation}
The equation (10) reduced to $\omega=-(1-\dot{\phi}^{2})$ for
spatially homogeneous field. Also remember that $\omega \neq
\omega_{1} + \omega_{2}$. The species of components of field are
recognize by constants $\alpha$ and $\beta$. If $\alpha$ and $\beta$
are both zero then one component is dust matter with zero EoS and
other is the cosmological constant (because of its EoS is
$\omega=-1$).\\
For simplicity and as a toy model we will assume that
$F(H)=K(H)=H^{2}$. Alternative models of dark energy suggest a
dynamical form of dark energy, which at least in an effective level,
can originate from a variable cosmological constant [24, 25], or
from various fields, such as a canonical scalar field [26-30]
(quintessence), a phantom field [31-38] or the quintom [39-51]. By
using some basic of quantum gravitational principles,
 we can formulate several other models for dark energy and in literature they are known as holographic dark energy paradigm [52-63]
 and agegraphic dark energy models [64-66].\\
Motivation to consider such splitting is to connect to the attempts
of unification of inflation, dark energy and dark matter appearing
in literature, which teach us that we can have fluids with exotic
EoS, we can consider modified gravity, which already can be
accounted as a base to consider fluids with exotic EoS [67-75].
However, in this stage it is to early to conclude that we found a
key approach to the unification problem. At least more deep research
should be performed, different types of functions should be
considered: from linear to some exotic forms not only function on
Hubble parameter, but function of scale factor, Ricci scalar or
functions of field and potential. We hope that systematic research
in this direction, very soon will give desirable results which we
will report systematically. We do not exclude possibility that in
some cases obtained results will not have deep connection with
Universe. Cosmology itself being interdisciplinary research field
and a possibility for testing different results has many open
questions, which allowed researchers make some phenomenological
assumptions which from first sight have not physical bases, but at
the same time we observed that with new experimental data old
phenomenological assumptions become as a key ingredient of modern
Cosmology and found fundamental physical interpretation.
\section{Field equations}
Field equations that govern the dynamics of our Universe in frame
work of general relativity read as,
\begin{equation}\label{eq:fieldeq}
R^{ij}-\frac{1}{2}g^{ij}R^{k}_{k}=T^{ij},
\end{equation}
where $T^{ij}$ is given by the equation (2). We consider the flat
FRW Universe with the following metric,
\begin{equation}\label{eq:FRW metric}
ds^{2}=dt^{2}-a(t)^{2}\left(\frac{dr^2}{1-kr^{2}}+r^{2}d\theta^{2}+r^{2}\sin^{2}\theta
d\phi^{2}\right).
\end{equation}
In that case the relation (11) gives the following
equations,
\begin{equation}\label{s13}
(\frac{\dot{a}}{a})^{2}=\frac{\rho}{3}-\frac{k}{a^{2}},
\end{equation}
and,
\begin{equation}\label{eq:Freidmann2}
- \frac{\ddot{a}}{a}=\frac{1}{6}(\rho+3P+\frac{3k}{a^{2}}).
\end{equation}
Also Bianchi identities implying that,
\begin{equation}\label{eq:Bieq}
\dot{\rho}+3H(\rho+P)=0,
\end{equation}
where $8\pi G=c=1$ and $\Lambda=0$ assumed. In this paper we
consider flat Universe with $k=0$. The interaction between
components formally splits the equation (15) into two equations,
with assumptions the transfer of energy from second component to
first component, which are given as,
\begin{equation}\label{eq:Bieq1}
\dot{\rho}_{1}+3H(\rho_{1}+P_{1})=Q,
\end{equation}
and
\begin{equation}\label{eq:Bieq1}
\dot{\rho}_{2}+3H(\rho_{2}+P_{2})=-Q.
\end{equation}
We take interaction strength $Q$ as phenomenological approach,
\begin{equation}\label{eq:interction}
Q=3bH\rho + \gamma \dot{\rho}
\end{equation}
where $b$ and $\gamma$ are constants, bound by observations and
$\rho=\rho_{1}+\rho_{2}$, $\dot{\rho}=\dot{\rho}_{1}+\dot{\rho}_{2}$
and $H=\dot{a}/a$ is Hubble parameter. By using this form of
interaction strength, the conservation of energy equations are
written as,
\begin{equation}\label{eq:Bieq1}
(1-\gamma)\dot{\rho}_{1}+3H\left(1+\omega_{1} - b -
b\frac{\rho_{2}}{\rho_{1}}\right)\rho_{1}=\gamma\dot{\rho}_{2},
\end{equation}
and
\begin{equation}\label{eq:Bieq1}
(1+\gamma)\dot{\rho}_{2}+3H\left(1+\omega_{2}+ b +
b\frac{\rho_{1}}{\rho_{2}}\right)\rho_{2}=-\gamma\dot{\rho}_{1}.
\end{equation}
Questions concerning to interacting dark energies are requiring
careful approach, because considered forms and types of interactions
between components are of phenomenological character. Different
interacting models of dark energy have been investigated. As far as
we know no known symmetry in nature prevents or suppresses a
non-minimal coupling between dark energy and dark matter. At the
same time, no piece of evidence has been so far presented against
interactions. It is found that a suitable interaction can help to
alleviate the coincidence problem. The forms of interaction term
considered in literature very often are of the following forms:
$Q=3Hb\rho_{dm}$, $Q=3Hb\rho_{\small{de}}$, or
$Q=3Hb\rho_{\small{tot}}$, where $b$ is a coupling constant and
positive $b$ means that dark energy decays into dark matter, while
negative $b$ means dark matter decays into dark energy. It was found
that the observations may favor the decaying of dark matter into
dark energy, while the second law of thermodynamics strongly favors
dark energy decays into dark matter.  Other forms for interaction
term considered in literature are $Q=\gamma\dot{\rho}_{\small{dm}}$,
$Q=\gamma\dot{\rho}_{\small{de}}$,
$Q=\gamma\dot{\rho}_{\small{tot}}$, $Q=3Hb
\rho_{i}+\gamma\dot{\rho_{i}}$, where $i=\{dm,de,tot\}$. These kind
of interactions are either positive or negative and can not change
sign. A sign-changeable interaction [76, 77] could be achieved very
simply either considering a possibility of including deceleration
parameter,
\begin{equation}\label{eq:signcinteraction}
Q=q( \gamma \dot{\rho}+3b H\rho),
\end{equation}
where $\alpha$ and $\beta$ are dimensionless constants, the energy
density $\rho$ could be $\rho_{m}$, $\rho_{\small{de}} $,
$\rho_{tot}$. $q$ is the deceleration parameter given by,
\begin{equation}\label{eq:decparameter}
q=-\frac{1}{H^{2}} \frac{\ddot{a}}{a}=-1-\frac{\dot{H}}{H^{2}}.
\end{equation}
The term of $\gamma \dot{\rho}$ in $Q$ is introduced from the
dimensional point of view. In the next section we investigate GSL of
thermodynamics in our system.
\section{The generalized second law of thermodynamics}
In this section we are going to deal with the question of the
validity of the GSL of thermodynamics. For GSL of thermodynamics we
will follow Ref. [78], where was considered validity of the GSL of
thermodynamics for the Universe bounded by the Hubble
horizon\footnote{ Recall that in case when $k=0$ as in our case
apparent horizon $R_{A}=\frac{1}{\sqrt{H^{2}+\frac{k}{a^{2}}}}$ we
get the radius of the Hubble horizon (\ref{eq:Habblehor}).},
\begin{equation}\label{eq:Habblehor}
R_{H}=\frac{1}{H},
\end{equation}
cosmological event horizon,
\begin{equation}\label{eq:cosevhor}
R_{E}= a\int_{t}^{\infty}\frac{dt}{a},
\end{equation}
and the particle horizon,
\begin{equation}\label{eq:particlehor}
R_{P}=a\int_{0}^{t}\frac{dt}{a}.
\end{equation}
The contents in the Universe bounded by the event horizons taken as
interacting two components of a single scalar field. The foundation
of GSL required the following Gibbs equation of thermodynamics
satisfied,
\begin{equation}\label{Gibs}
T_{X}dS_{IX}= PdV_{X} + dE_{IX}
\end{equation}
where $S_{IX}$ and $E_{IX}=\rho V_{X}$, are internal entropy and
energy within the horizon respectively, while $V_{X}=\frac{4}{3}\pi
R^{3}_{X}$ denotes the volume of sphere with the horizon radius
$R_{A}$. Recall that GSL together with the first law of
thermodynamics for the time derivative of total entropy gives,
\begin{equation}\label{eq:UF}
\dot{S}_{X}+\dot{S}_{IX}=\frac{R^{2}_{X}}{GT_{X}}\left(\frac{k}{a^{2}}-\dot{H}\right)\dot{R}_{X},
\end{equation}
while in the case of without the first law we get,
\begin{equation}\label{eq:UNUF}
\dot{S}_{X}+\dot{S}_{IX}= \frac{2\pi R_{X}}{G}\left[ R^{2}_{X}\left(\frac{k}{a^{2}}-\dot{H}\right)(\dot{R}_{X}-HR_{X})+\dot{R}_{X} \right].
\end{equation}
Under the notations used above we understood that
$T_{X}=\frac{1}{2\pi R_{X} }$ and $R_{X}$ is temperature and radius
for a given horizon under equilibrium thermodynamics respectively,
$S_{X}$ is the horizon entropy and $\dot{S}_{IX}$ as the rate of
change of internal entropy. It was found that the first and second
laws of thermodynamics hold on the apparent horizon when the
apparent horizon and the event horizon of the Universe are
different, while for consideration of only event horizon these laws
breakdown [79]. The Friedmann equations and the first law of
thermodynamics (on the apparent horizon) are equivalent if the
Universe is bounded by the apparent horizon $R_{A}$ with temperature
$T_{A}=\frac{1}{2\pi R_{A}}$ and entropy $S_{A}=\frac{\pi
R_{A}}{G}$. Usually, the Universe bounded by apparent horizon and in
this region the Bekenstein's entropy-mass bound ($S \leq 2\pi E
R_{A}$) and entropy - area bound ($S \leq\frac{A}{4}$) are hold.
Numerical analysis of the equations (27) and (28) given in the
section 6. In order the GSL to be hold it is required that
$\dot{S}_{X}+\dot{S}_{IX}\geq0$ i.e. the sum of entropy of matter
enclosed by horizon must be not be a decreasing function of time.
\section{Statefinder diagnostics}
In this section we consider problem of statefinder diagnostics. The
property of dark energy is model dependent and to differentiate
various models of dark energy, a sensitive diagnostic tool is
needed. Hubble parameter $H$ and deceleration parameter $q$ are very
important quantities which can describe the geometric properties of
the Universe. Since $\dot{a}>0$, hence $H>0$ means the expansion of
the Universe. Also, $\ddot{a}>0$, which is $q<0$ indicates the
accelerated expansion of the universe. Since, the various dark
energy models give $H>0$ and $q<0$, they can not provide enough
evidence to differentiate the more accurate cosmological
observational data and the more general models of dark energy. For
this aim we need  higher order of time derivative of scale factor
and geometrical tool. Sahni \emph{et.al} [79] proposed geometrical
statefinder diagnostic tool, based on dimensionless parameters $(r,
s)$ which are function of scale factor and its time derivative.
These are jerk and snap parameters which are define as the
following,
\begin{eqnarray}\label{eq:statefinder}
r&=&\frac{1}{H^{3}}\frac{\dddot{a}}{a},\nonumber\\
s&=&\frac{r-1}{3(q-\frac{1}{2})},
\end{eqnarray}
respectively. The deceleration parameter is also given by the
equation (22). We will use other form of parameters in terms of he
total energy density $\rho$ and pressure $P$ in the Universe,
\begin{eqnarray}\label{eq:rsrhop}
r&=&1+\frac{9(\rho+P)}{2\rho}\frac{\dot{P}}{\dot{\rho}},\nonumber\\
s&=&\frac{(\rho+P)}{P}\frac{\dot{P}}{\dot{\rho}},\nonumber\\
q&=&\frac{1}{2}(1+\frac{3P}{\rho}).
\end{eqnarray}
The plot of $r$, $s$ and $q$ are given in the Fig. 1.\\
\begin{figure}[h]
 \begin{center}$
 \begin{array}{cccc}
 \includegraphics[width=55 mm]{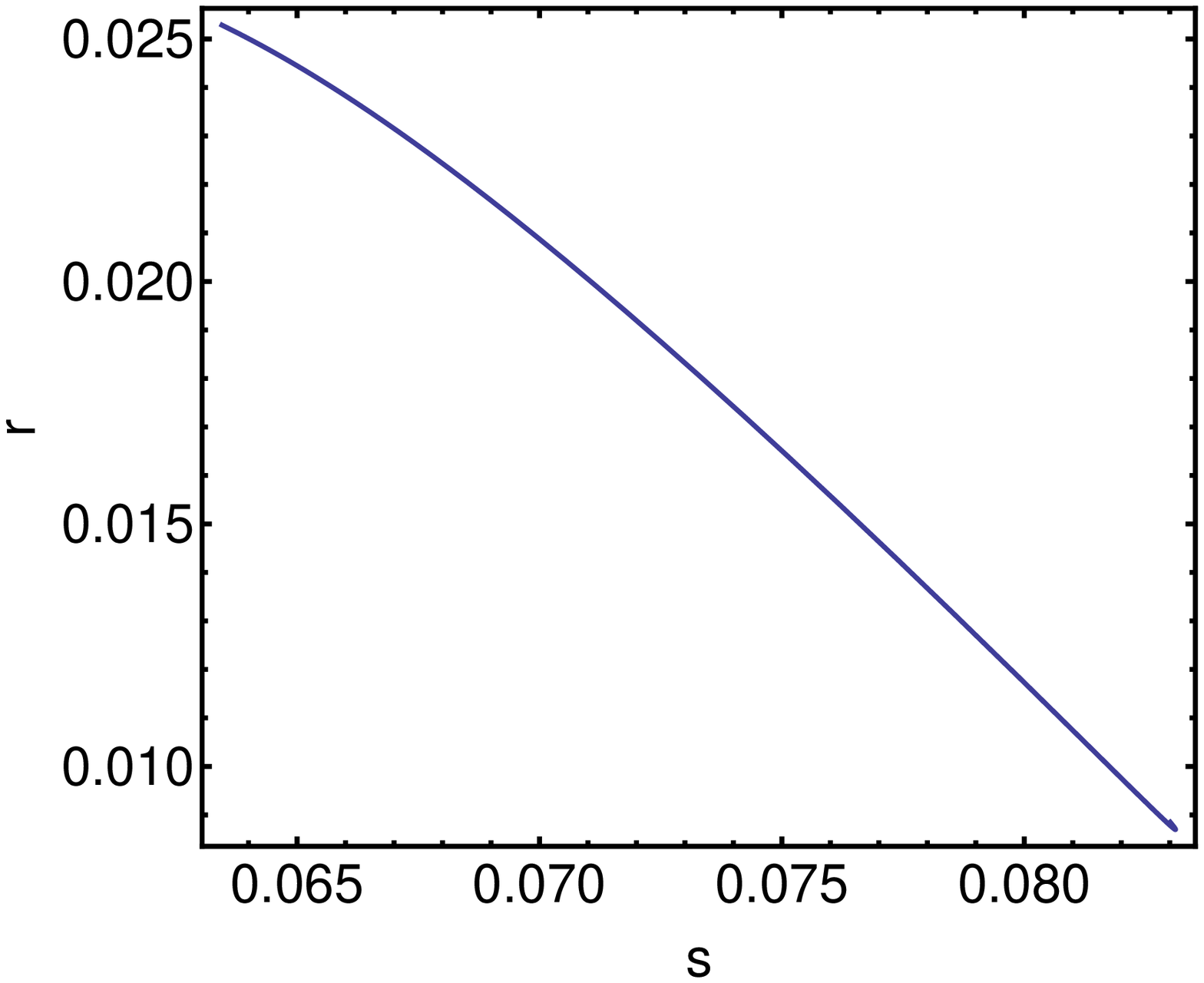}
 \includegraphics[width=55 mm]{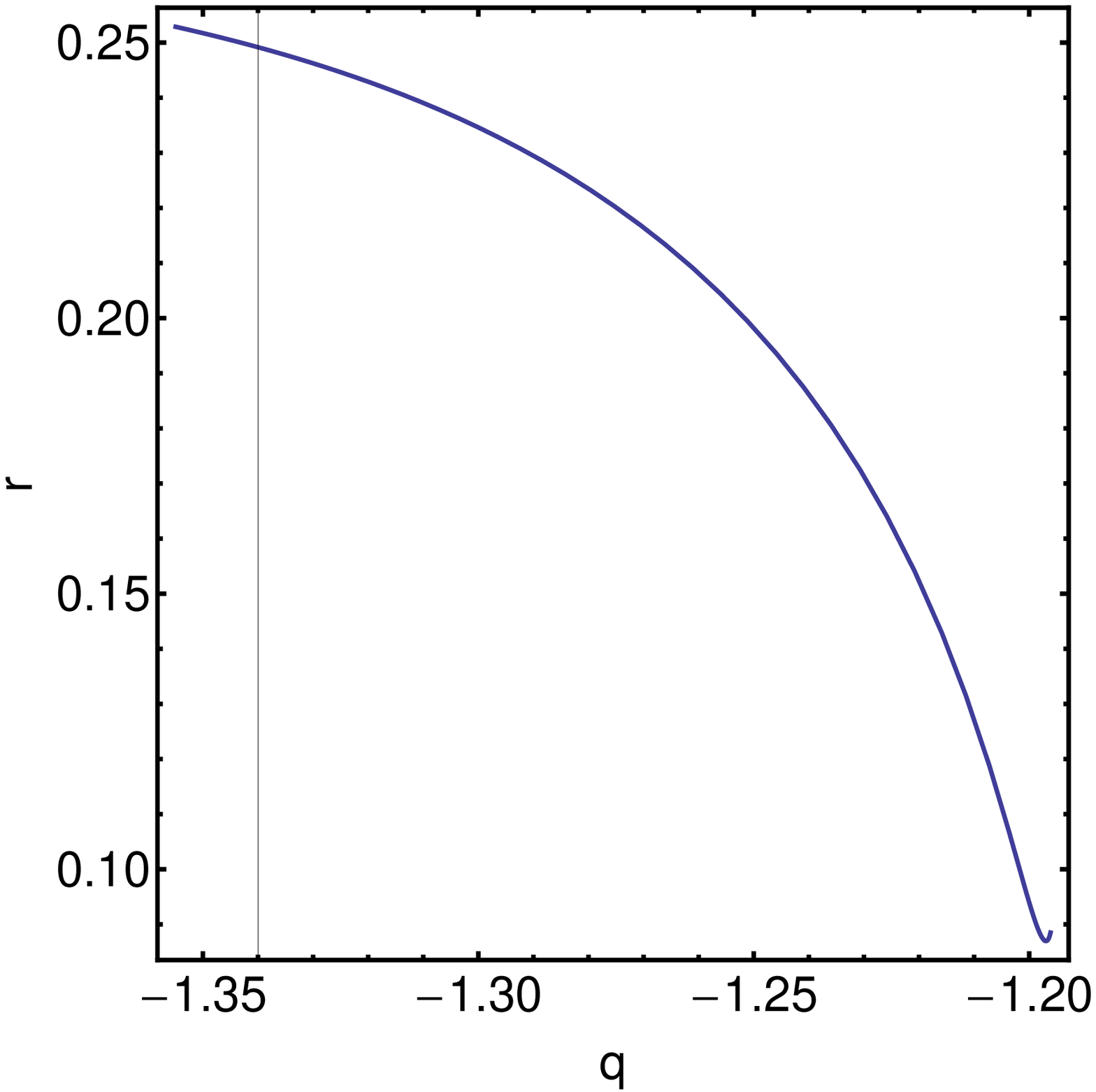}
 \includegraphics[width=55 mm]{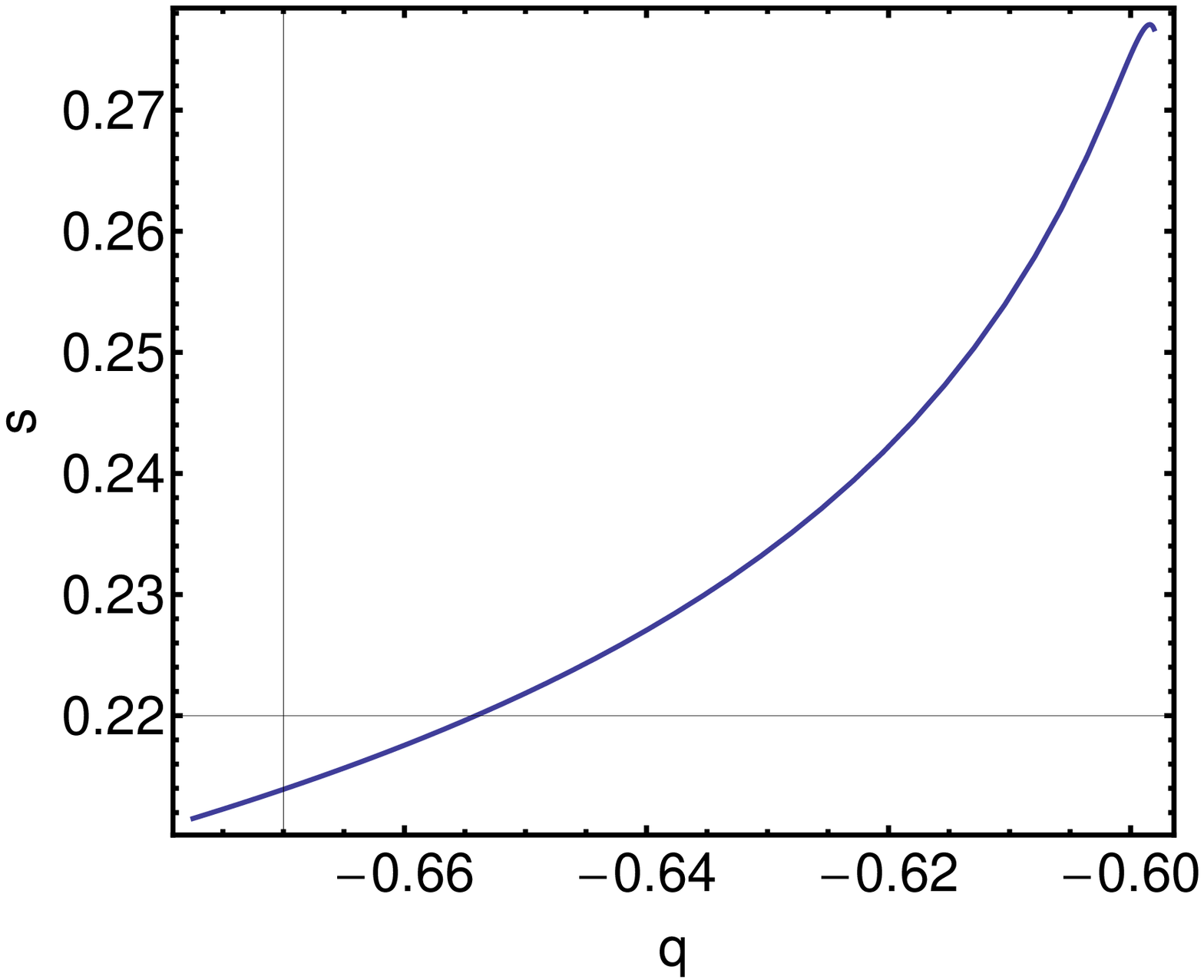}
 \end{array}$
 \end{center}
 \caption{The above plot refer to the links between $r$, $s$ and $q$. We choose $\alpha=2.5$, $\beta=1.5$, $b=0.01$ and $\gamma=0.02$. }
 \label{fig:1}
\end{figure}

\begin{figure}[h]
 \begin{center}$
 \begin{array}{cccc}
 \includegraphics[width=55 mm]{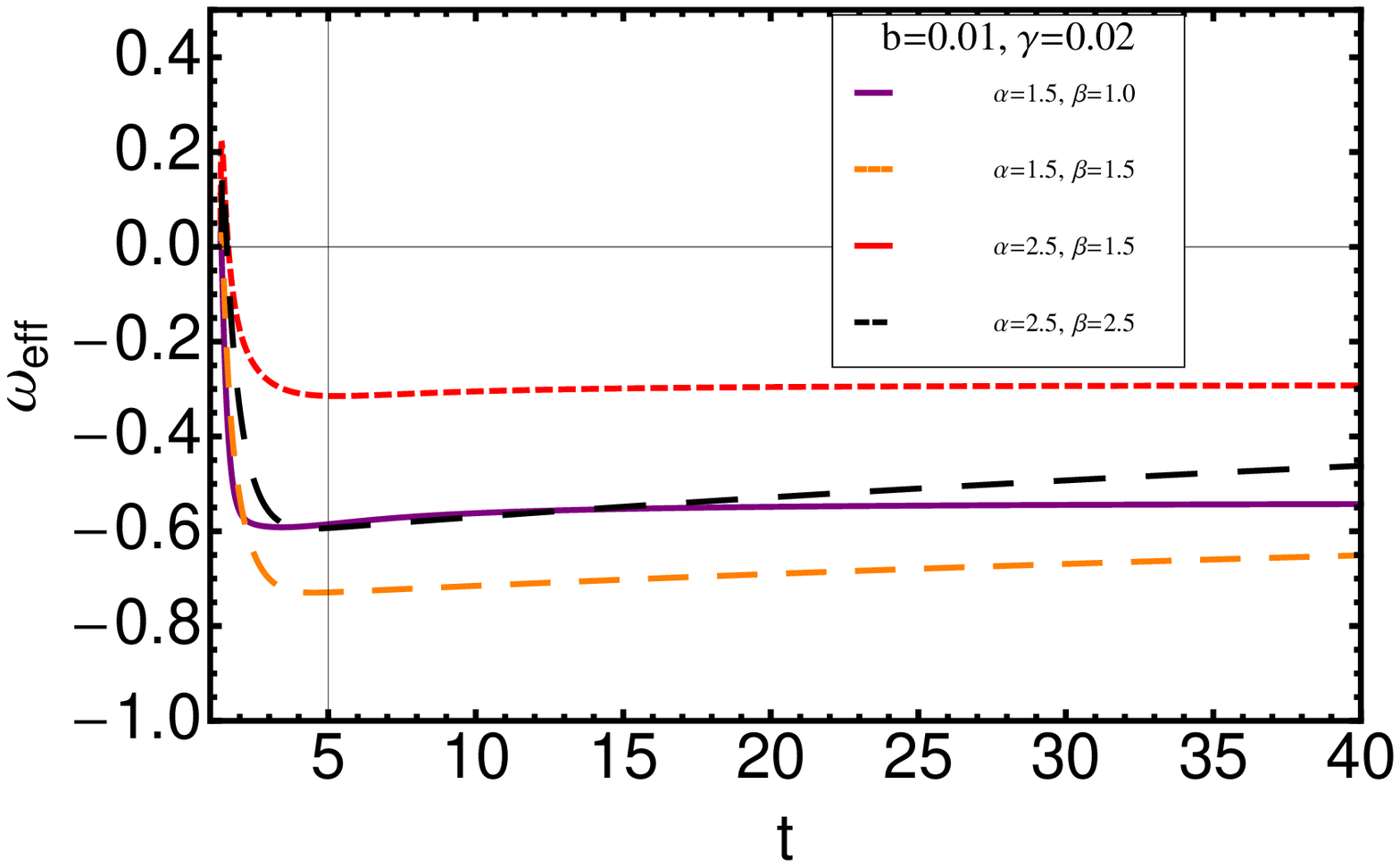} &
 \includegraphics[width=55 mm]{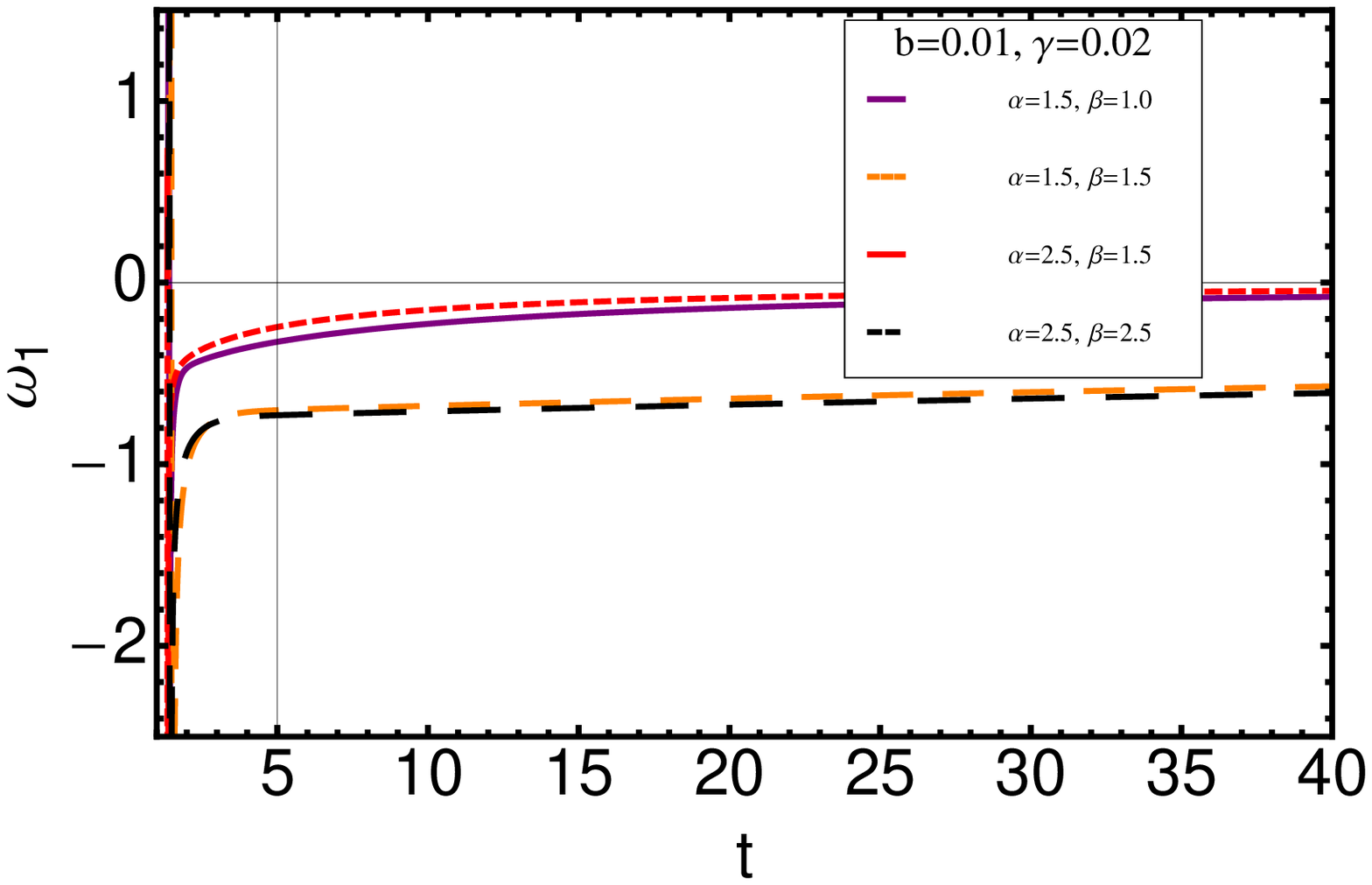} \\
 \includegraphics[width=55 mm]{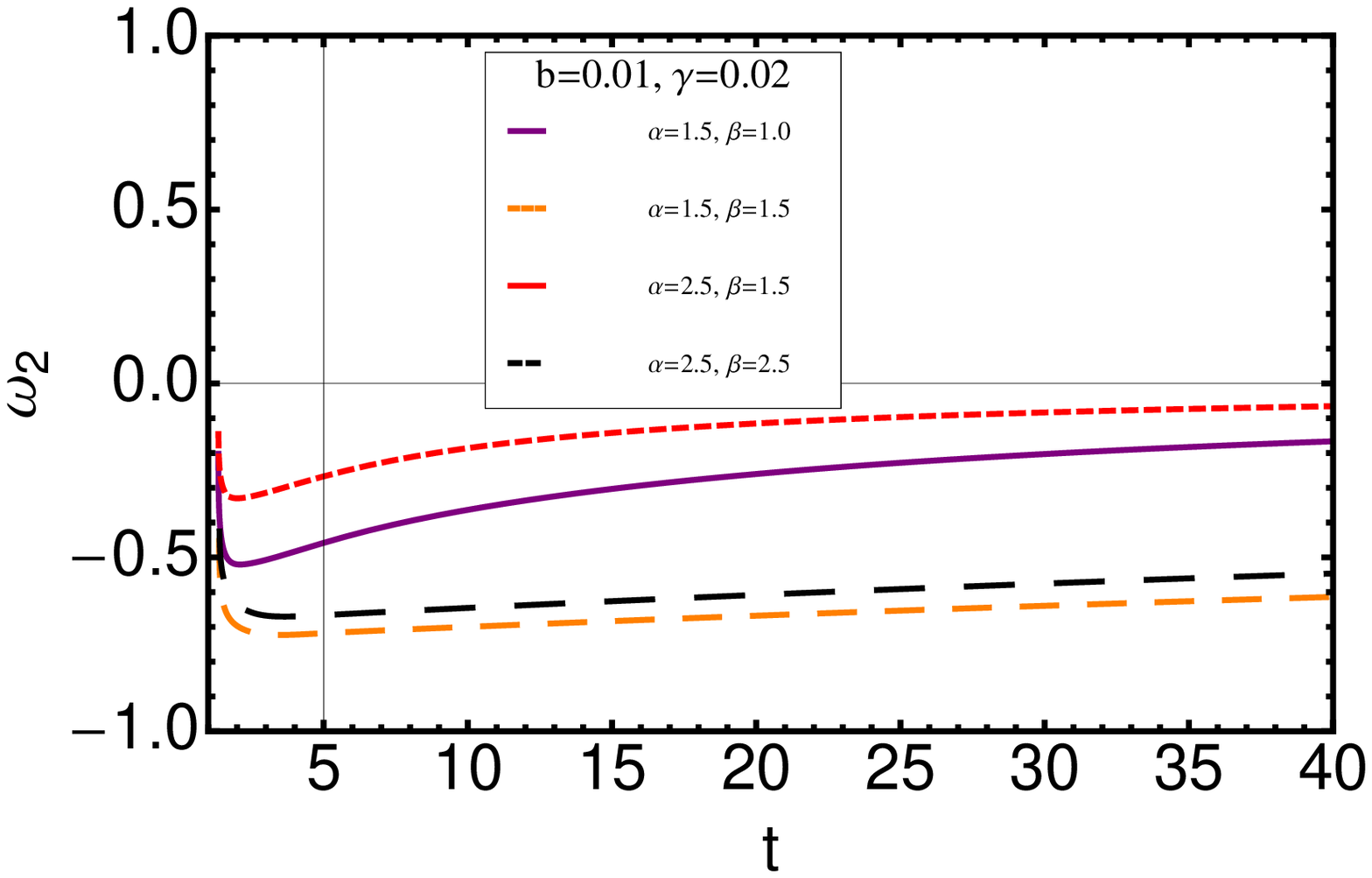} &
 \includegraphics[width=55 mm]{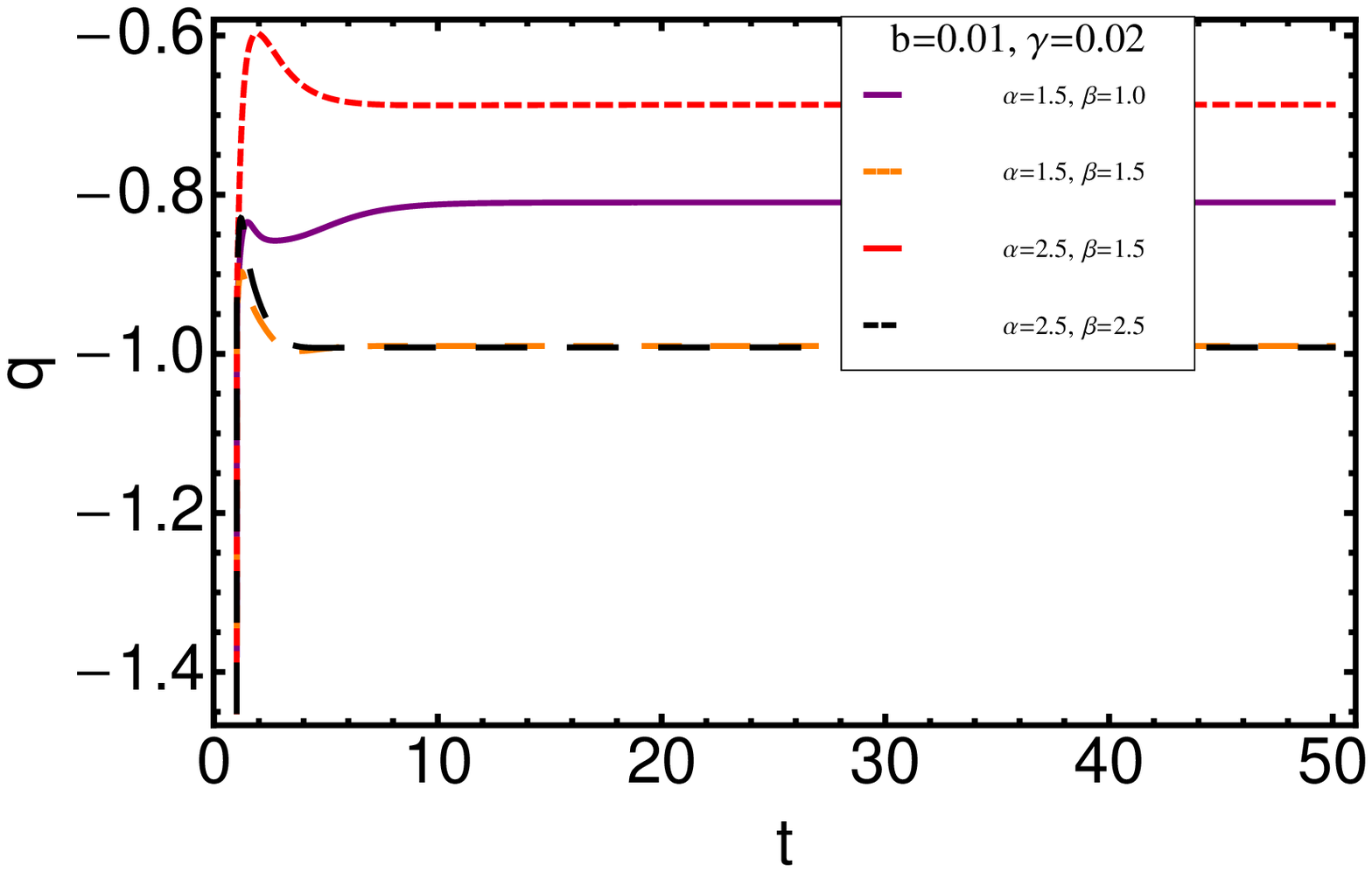}
 \end{array}$
 \end{center}
 \caption{The above plots refer to the links between $\omega_{tot}$, $\omega_{1}$, $\omega_{2}$, $q$ and $t$ for fixed interaction parameters.
 In $(q-t)$ plot, it is seen that the Universe initially undergoes a rapidly falling acceleration followed by a rise in it.
 At a particular epoch  the Universe get into a phase of constant acceleration in which we are presently located. We choose $b=0.01$ and $\gamma=0.02$.}
 \label{fig:2}
\end{figure}
\begin{figure}[h]
 \begin{center}$
 \begin{array}{cccc}
 \includegraphics[width=50 mm]{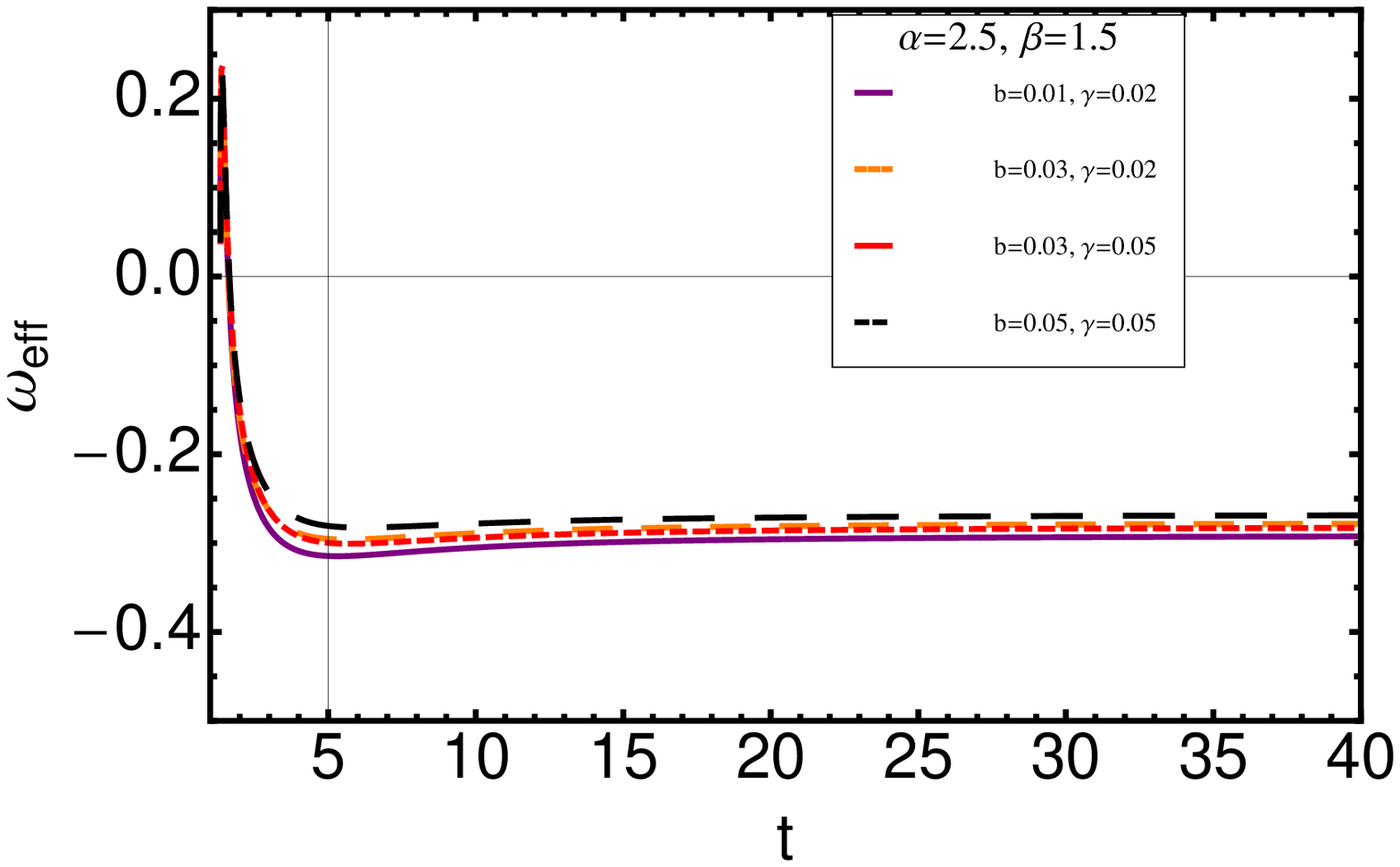} &
 \includegraphics[width=50 mm]{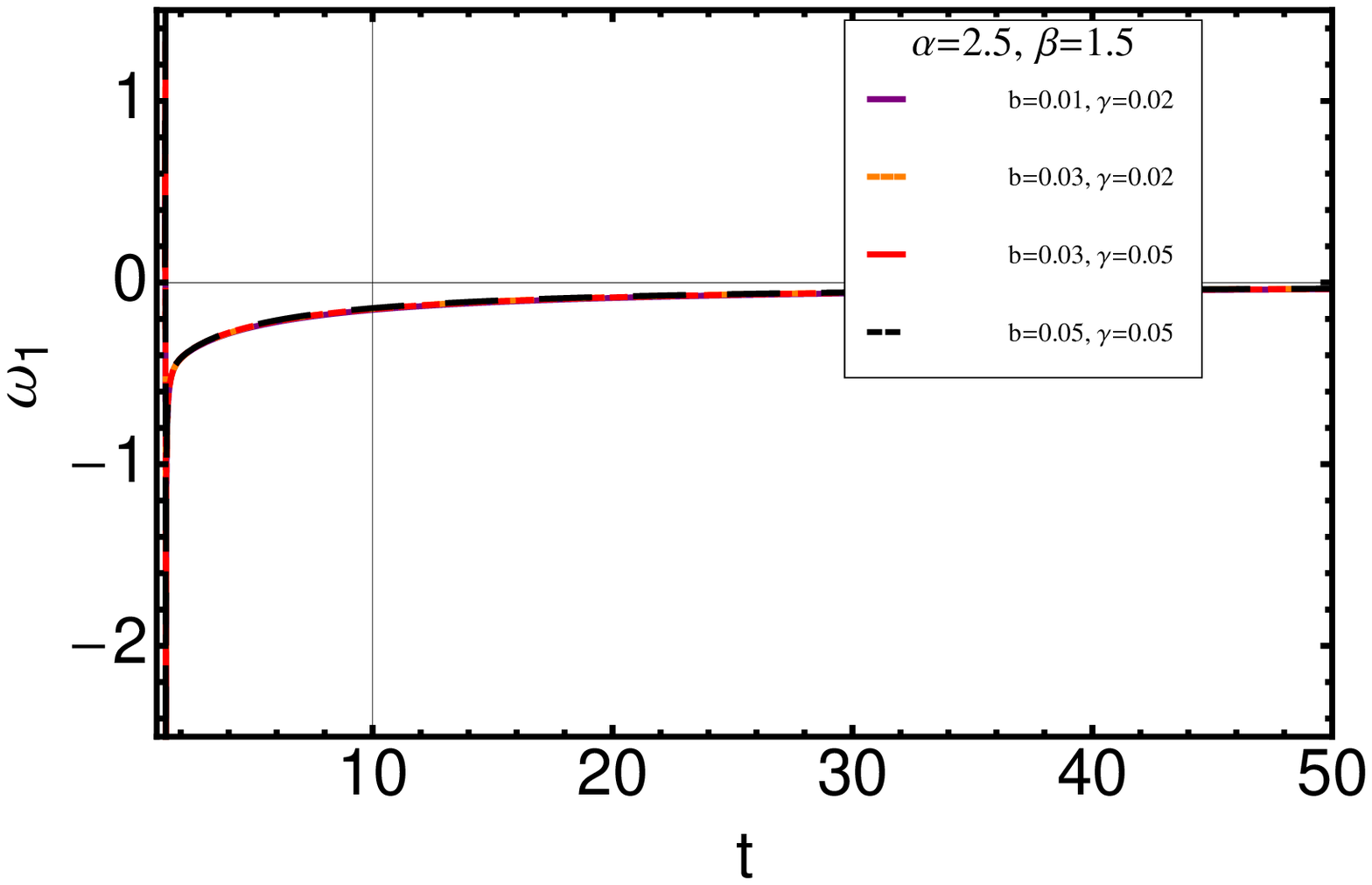} \\
 \includegraphics[width=50 mm]{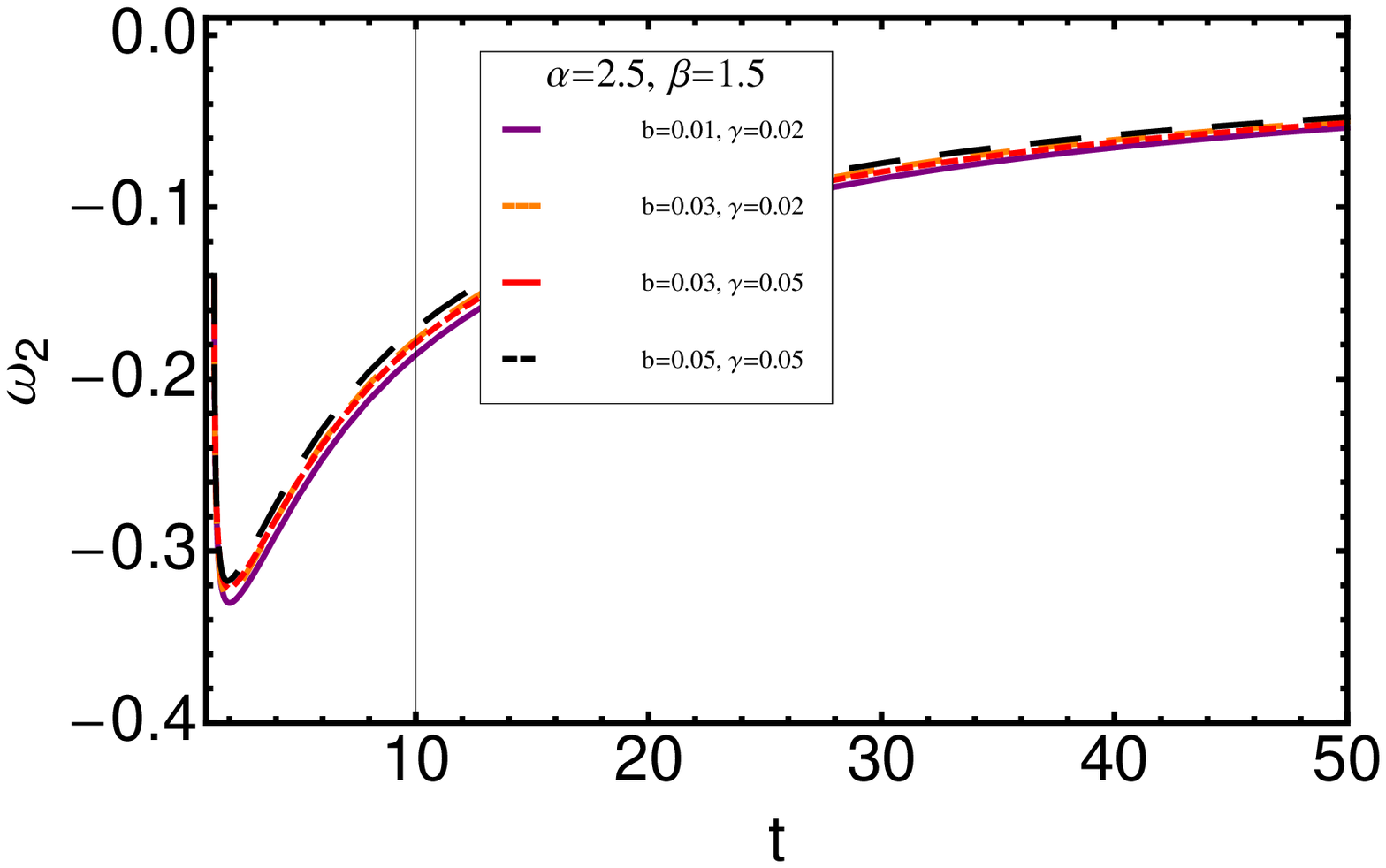} &
 \includegraphics[width=50 mm]{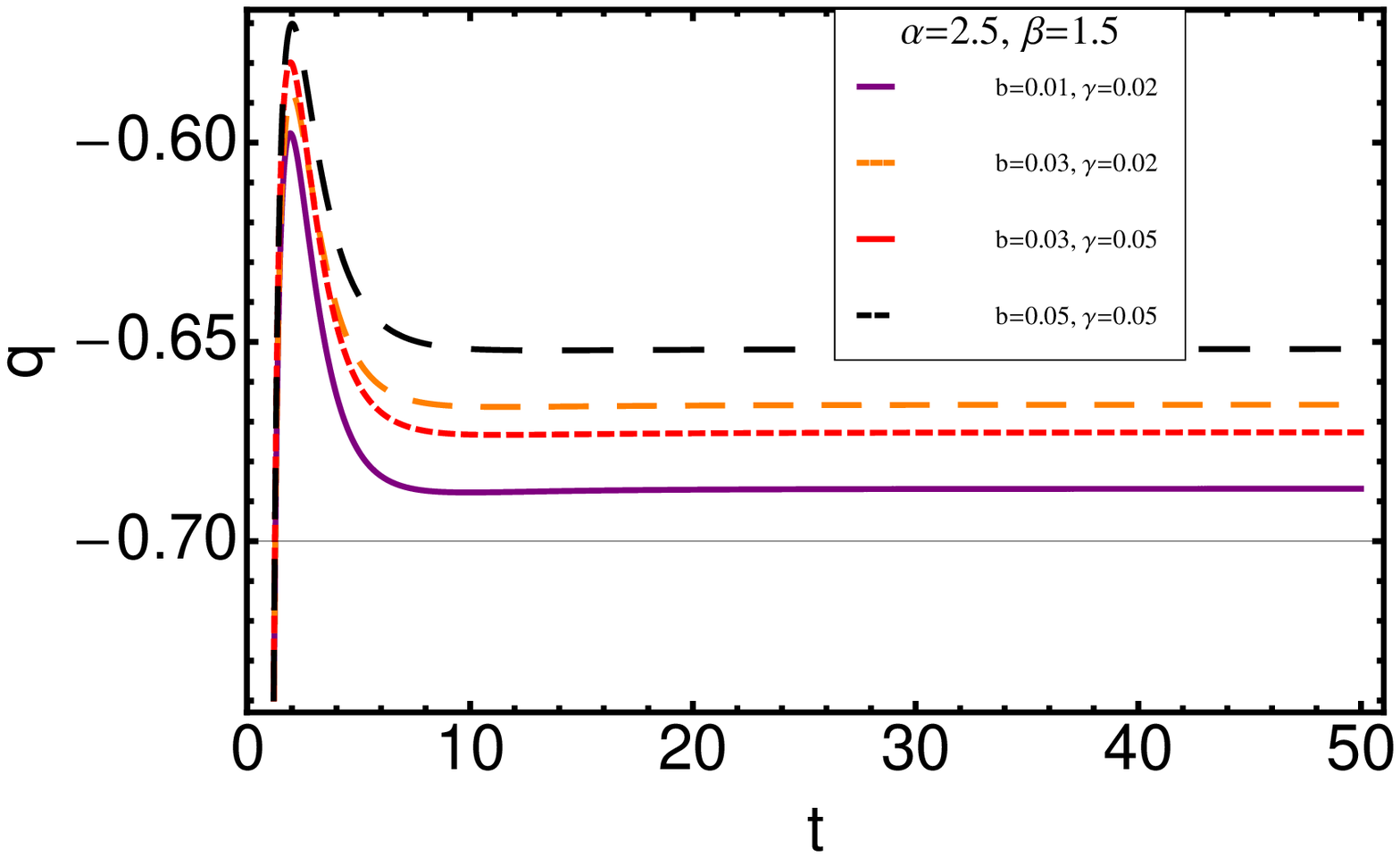}
 \end{array}$
 \end{center}
 \caption{The above plots refer to the links between $\omega_{tot}$, $\omega_{1}$, $\omega_{2}$, $q$ and $t$ for fixed fluid parameters.
 In $(q-t)$ plot, it is seen that the Universe initially undergoes a rapidly falling acceleration followed by a rise in it.
 At a particular epoch  the Universe get into a phase of constant acceleration in which we are presently located. We choose $\alpha=2.5$ and $\beta=1.5$.}
 \label{fig:3}
\end{figure}

\section{Numerical Results}
Numerical analysis provide us following information. First of all we
find that field is a complex one and in this case a complex field is
able to produce accelerated expansion i.e $q<0$. In the Fig. 2 we
present behavior of $\omega_{eff}=\frac{P_{1}+
P_{2}}{\rho_{1}+\rho_{2}}$ of composed fluid, $\omega_{1}$,
$\omega_{2}$ of fluids and $q$ against time $t$ for different values
of $\alpha$ and $\beta$ with the fixed valued of interacting
constants. We observe that for the composed fluid for early stages
of evolution it is a fluid with $\omega>0$ but then we have
transition to a dark energy with $\omega>-1$ indicating
quintessence-like behavior. We also investigate behavior of
$\omega_{1}$ and $\omega_{2}$ of our phenomenological fluids and
observe that $\omega_{1}$ is a positive at early stages and carries
fast jump to a dark energy with negative EoS parameter (phantom dark
energy) and then during evolution it becomes quintessence with
$\omega>-1$. Fast change of the type of the first fluid can have
very deep physics. In this letter we assume that, or this behavior
related to the fact that Tachyon field is unstable in early stages
of evolution, either that first fluid itself is not in
thermodynamical equilibrium and some irreversible processes like to
a particle creation and annihilation is happening "inside" the
fluid. More deep analysis is needed in order to conclude with right
physics. On the other hand, we see that $\omega_{2}$ is completely
negative for all values of our parameters ($0>\omega>-1$) which
suggest quintessence-like behavior. For the deceleration parameter
we see, that it is negative
i.e. we get ever accelerated Universe.\\
In the second analysis of the model, for illustration, we fixed
values of parameters coming from fluids splitting i.e. $\alpha$ and
$\beta$ and investigate behavior of EoS parameters and $q$ against
time for different values of interacting constants (see Fig. 3). We
conclude that, for different values of interaction parameters
intensively considered in literature discussed from experimental
constraints, composed fluid and its components can be identified as
the same and first fluid has the same behavior as already discussed
above i.e fast transition from a fluid to a dark energy at early
stages of
evolution. Again, we have ever accelerated Universe.\\
Validity of the GSL of thermodynamics is satisfied for our model,
which can be accounted as a simple and first test telling us about
true-like model (see Fig. 4). Other test like comparing the model
with experimental data can be also done in future, which is also
will allow us to fix a viable range of the values of the model.\\
Finally in the Fig. 5 we present $V(t)$ and $\phi(t)$. We can see
that there is a maximum for the potential which is corresponding to
the minimum of the $\omega_{eff}$. Then, Tachyon potential vanished
at the late time, which is expected. We find that increasing $\beta$
increased potential while increasing $\alpha$ decreased one. Also we
see that $\phi$ is increasing function of time at the late time,
while in the early stage it is strongly depend on choosing
parameters. We can observe that the parameter $\beta$ decreases
value of $\phi$ at the late stage but the parameter $\alpha$
increases one.
 \begin{figure}[h]
 \begin{center}$
 \begin{array}{cccc}
 \includegraphics[width=45 mm]{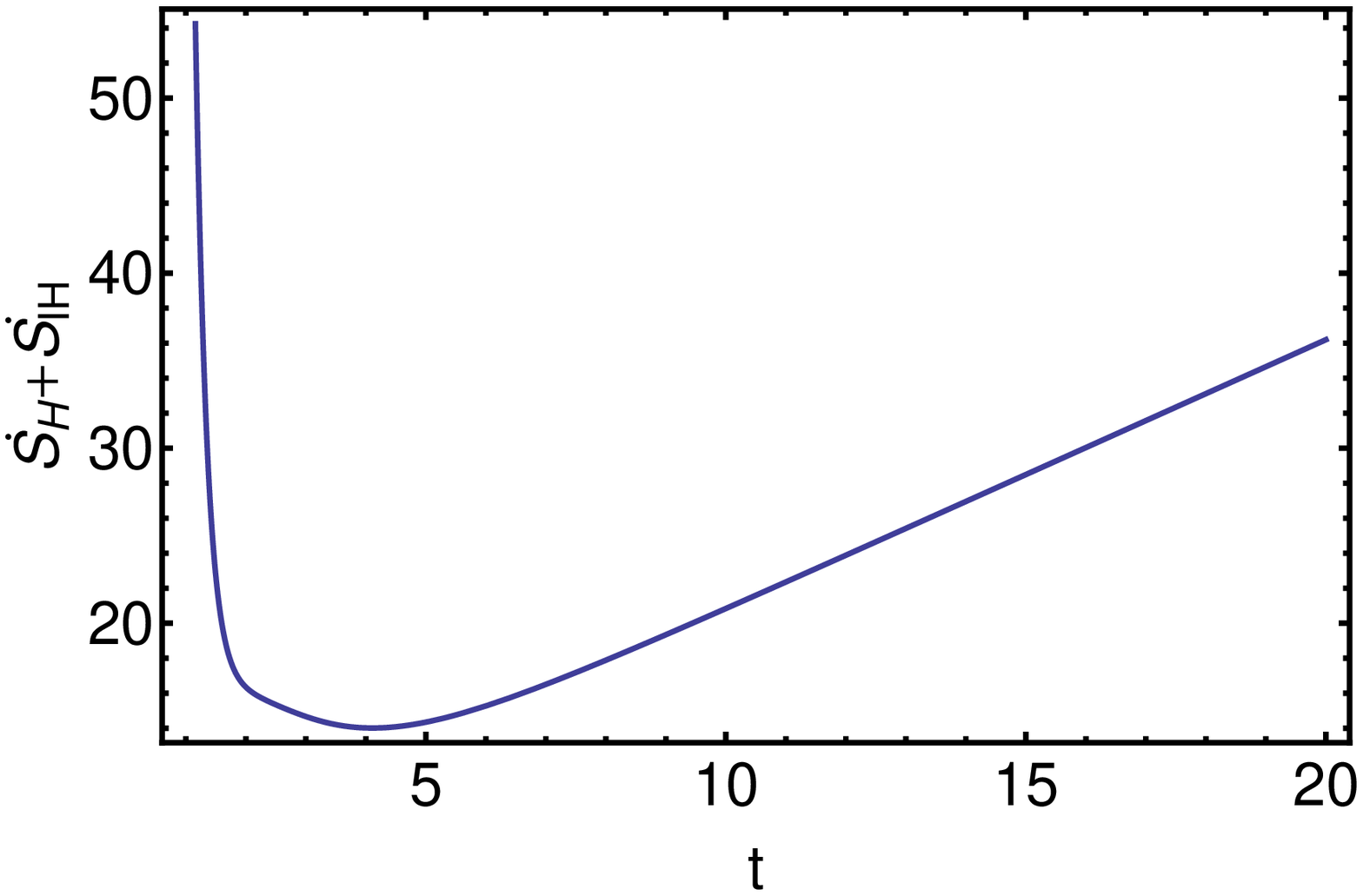} &
 \includegraphics[width=45 mm]{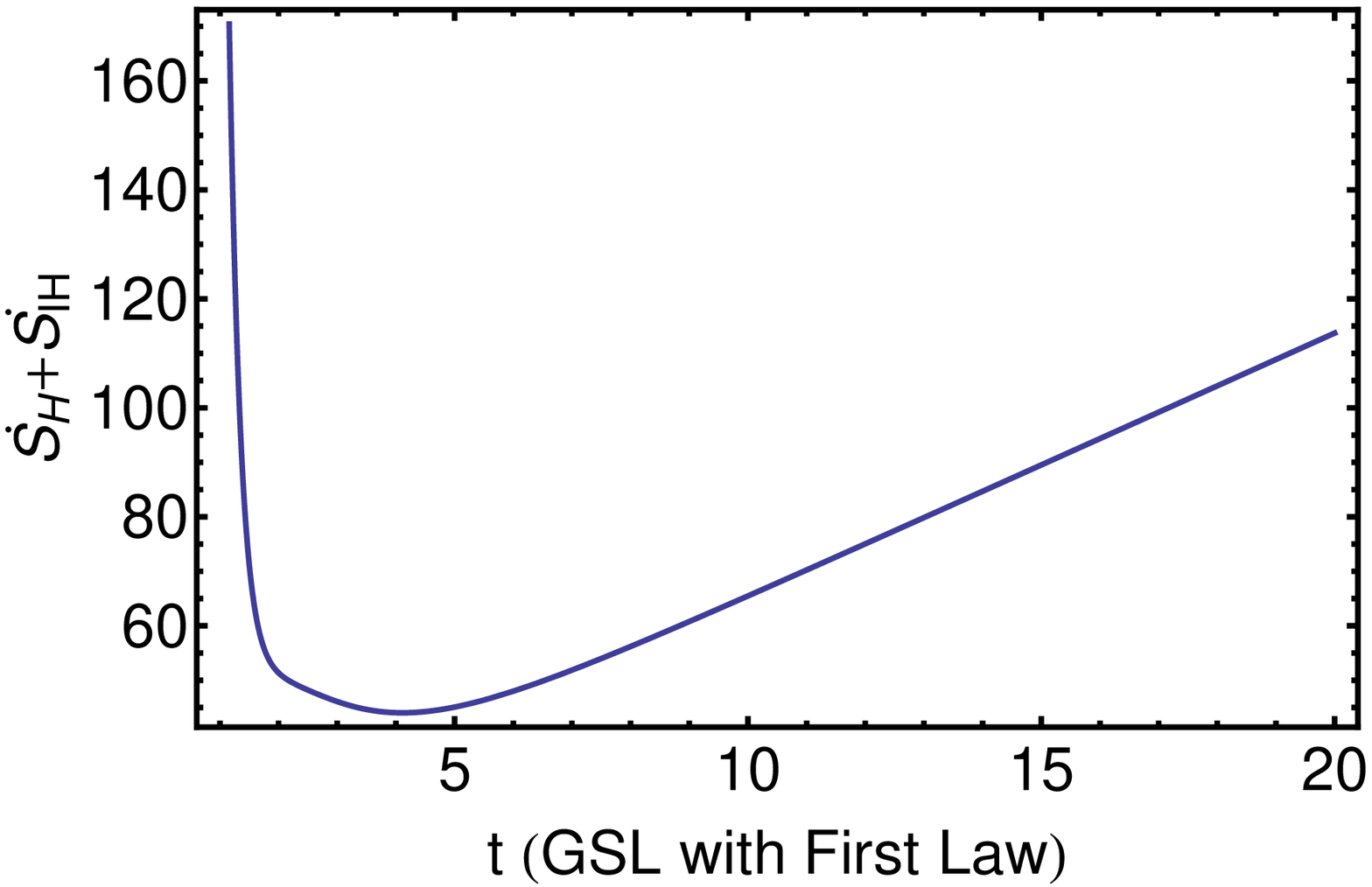} \\
 \end{array}$
 \end{center}
 \caption{$\dot{S}_{H}+\dot{S}_{IH}$ versus time for the case of, Left: only GSL of thermodynamics (without the first law).
 Right: GSL together with the first law of thermodynamics. We set $\alpha=2.5$, $\beta=1.5$, $b=0.01$ and $\gamma=0.02$. }
 \label{fig:4}
\end{figure}

\begin{figure}[h]
 \begin{center}$
 \begin{array}{cccc}
 \includegraphics[width=60 mm]{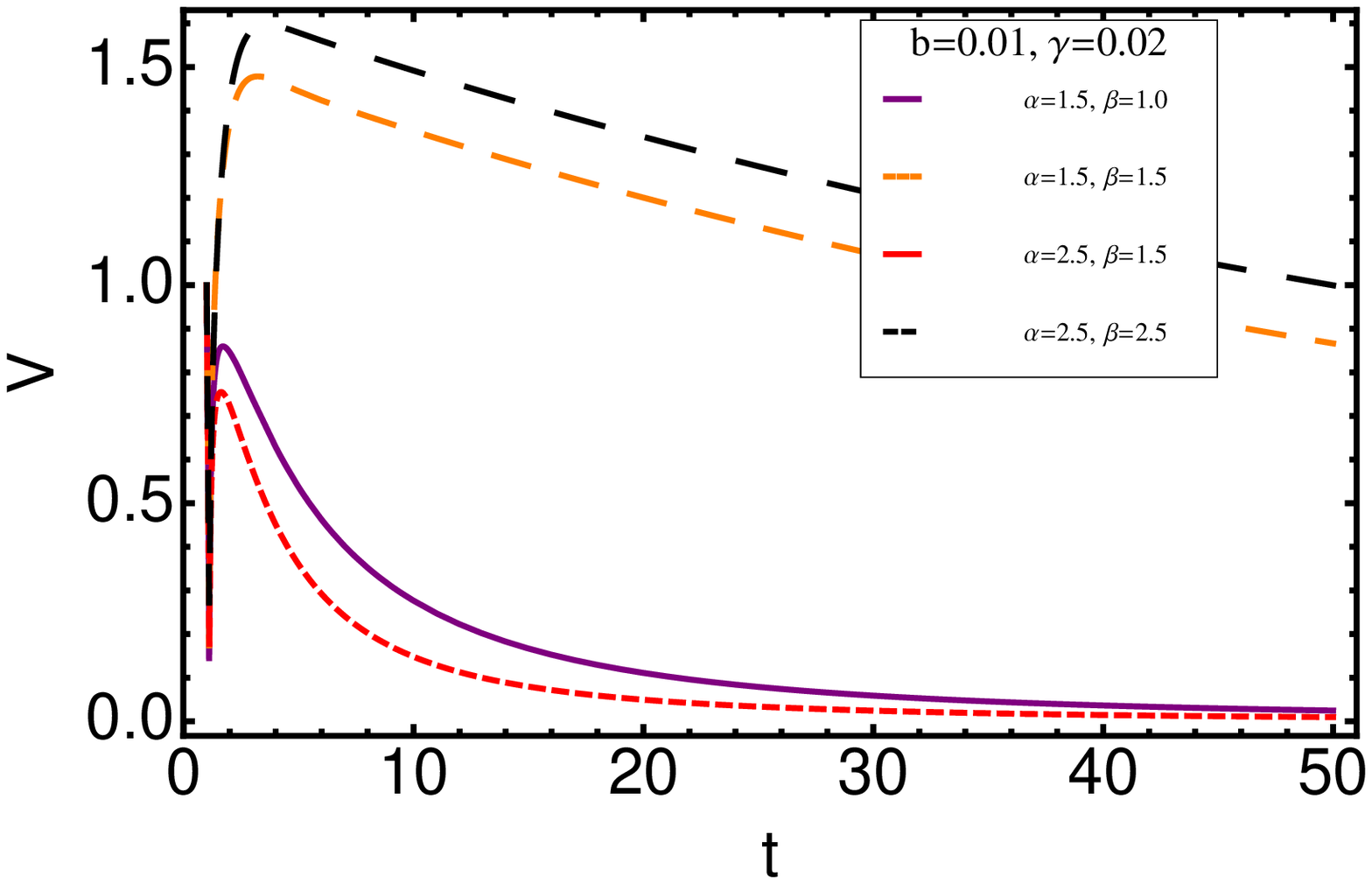} &
 \includegraphics[width=60 mm]{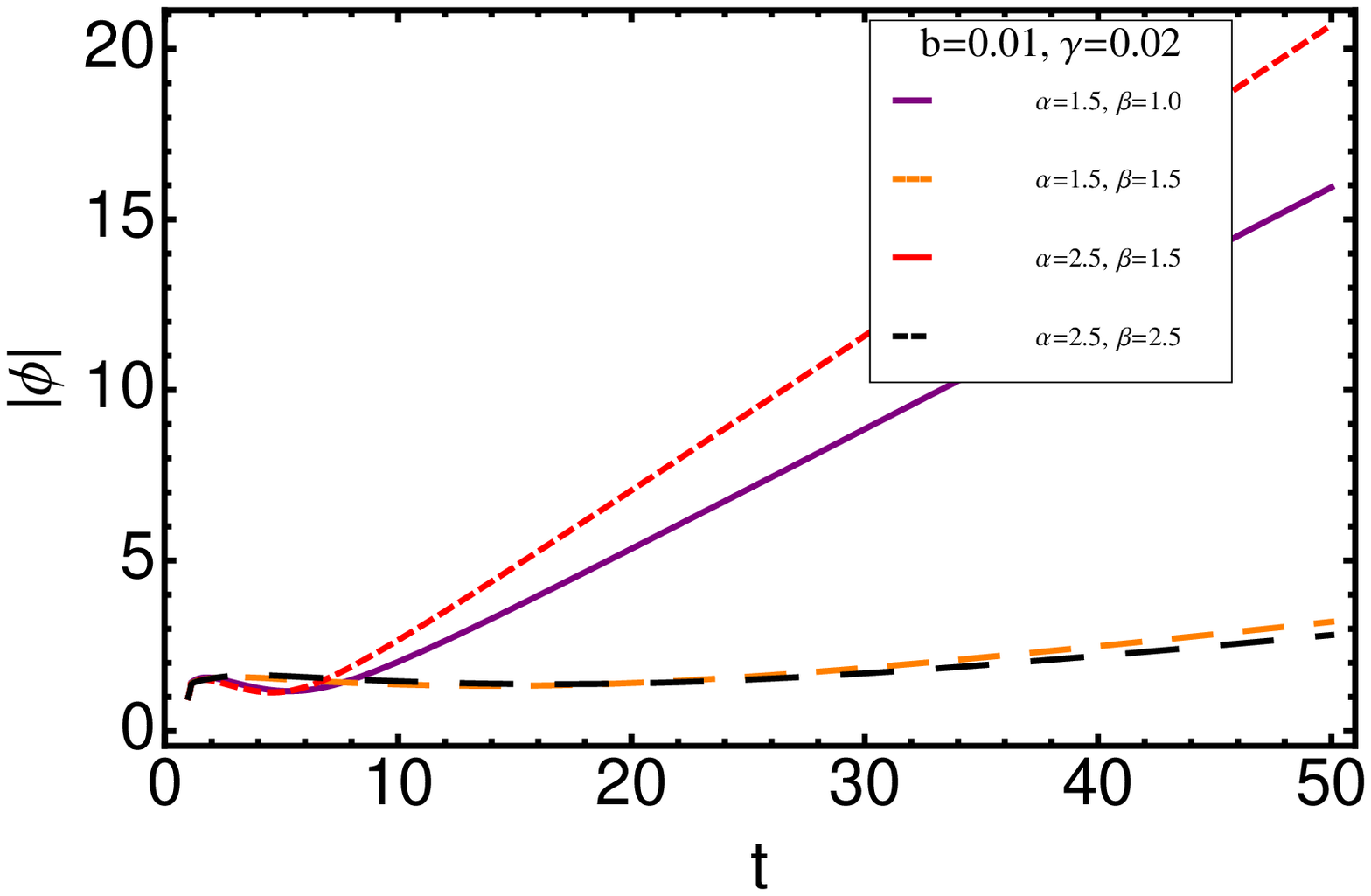} &
 \end{array}$
 \end{center}
 \caption{Plots of $V(t)$ and $\phi(t)$, where we choose $b=0.01$ and $\gamma=0.02$. }
 \label{fig:5}
\end{figure}
\section{Conclusion}
In this article we considered a phenomenological splitting of
Tachyonic scalar field which gives rise of two fluids. In base of
special form of additional term, which is a function of Hubble
parameter with an interaction between them we investigated behavior
of the Universe. We observed that we have ever accelerated
expansion. Fluid responsible for that acceleration for later stages
is dark energy with $\omega<0$, while for early stages of evolution
it was not dark energy, moreover we have fluid type transition. This
behavior can be accepted as counterintuitive related to the fact
that, in GR, accelerated expansion caused by dark energy. GR does
not give any explanation related to the origin of dark energy as
well as about origin of dark meter. Numerical analysis shows that
field is a complex and is able to provide discussed acceleration.
Validity of the GSL of thermodynamics could be thought as a test
giving us a hope, that we could continue our research in this
direction. We already mentioned about motivations and about our
interests related to the question, which is arose from the research
providing different ways for unification of Inflation, dark energy
and dark matter, which provides us knowledge that we have a right to
consider fluids with exotic EoS like in our case, with one
difference that the base of our fluids is Tachyonic field. In future
we would like to consider different variations concerning to the
forms of functions and parameters and investigate behavior of the
Universe and fluid. Also, one can investigate the effects of shear
or bulk viscosities on the system.
\section*{Acknowledgments}
Martiros Khurshudyan has been supported by EU fonds in the frame of the program FP7-Marie Curie Initial Training Network INDEX NO.289968.

\end{document}